\PassOptionsToPackage{unicode}{hyperref}
\PassOptionsToPackage{hyphens}{url}
\PassOptionsToPackage{dvipsnames,svgnames,x11names}{xcolor}
\documentclass[
  11pt]{article}
\usepackage{xcolor}
\usepackage[margin=1in]{geometry}
\usepackage{amsmath,amssymb}
\usepackage{iftex}
\ifPDFTeX
  \usepackage[T1]{fontenc}
  \usepackage[utf8]{inputenc}
  \usepackage{textcomp} 
\else 
  \usepackage{unicode-math} 
  \defaultfontfeatures{Scale=MatchLowercase}
  \defaultfontfeatures[\rmfamily]{Ligatures=TeX,Scale=1}
\fi
\usepackage{lmodern}
\ifPDFTeX\else
\fi
\IfFileExists{upquote.sty}{\usepackage{upquote}}{}
\IfFileExists{microtype.sty}{
  \usepackage[]{microtype}
  \UseMicrotypeSet[protrusion]{basicmath} 
}{}
\makeatletter
\@ifundefined{KOMAClassName}{
  \IfFileExists{parskip.sty}{%
    \usepackage{parskip}
  }{
    \setlength{\parindent}{0pt}
    \setlength{\parskip}{6pt plus 2pt minus 1pt}}
}{
  \KOMAoptions{parskip=half}}
\makeatother
\makeatletter
\ifx\paragraph\undefined\else
  \let\oldparagraph\paragraph
  \renewcommand{\paragraph}{
    \@ifstar
      \xxxParagraphStar
      \xxxParagraphNoStar
  }
  \newcommand{\xxxParagraphStar}[1]{\oldparagraph*{#1}\mbox{}}
  \newcommand{\xxxParagraphNoStar}[1]{\oldparagraph{#1}\mbox{}}
\fi
\ifx\subparagraph\undefined\else
  \let\oldsubparagraph\subparagraph
  \renewcommand{\subparagraph}{
    \@ifstar
      \xxxSubParagraphStar
      \xxxSubParagraphNoStar
  }
  \newcommand{\xxxSubParagraphStar}[1]{\oldsubparagraph*{#1}\mbox{}}
  \newcommand{\xxxSubParagraphNoStar}[1]{\oldsubparagraph{#1}\mbox{}}
\fi
\makeatother

\usepackage{longtable,booktabs,array}
\usepackage{calc} 
\usepackage{etoolbox}
\makeatletter
\patchcmd\longtable{\par}{\if@noskipsec\mbox{}\fi\par}{}{}
\makeatother
\IfFileExists{footnotehyper.sty}{\usepackage{footnotehyper}}{\usepackage{footnote}}
\makesavenoteenv{longtable}
\usepackage{graphicx}
\makeatletter
\newsavebox\pandoc@box
\newcommand*\pandocbounded[1]{
  \sbox\pandoc@box{#1}%
  \Gscale@div\@tempa{\textheight}{\dimexpr\ht\pandoc@box+\dp\pandoc@box\relax}%
  \Gscale@div\@tempb{\linewidth}{\wd\pandoc@box}%
  \ifdim\@tempb\p@<\@tempa\p@\let\@tempa\@tempb\fi
  \ifdim\@tempa\p@<\p@\scalebox{\@tempa}{\usebox\pandoc@box}%
  \else\usebox{\pandoc@box}%
  \fi%
}
\def\fps@figure{htbp}
\makeatother

\usepackage{amsmath,amssymb,amsthm}

\usepackage{booktabs}
\usepackage{graphicx}
\usepackage{natbib}
\usepackage{hyperref}
\usepackage{xcolor}
\usepackage[section]{placeins}  
\usepackage{siunitx}            
\usepackage{setspace}
\usepackage{caption}
\usepackage{tikz}\usetikzlibrary{angles,quotes,calc}  
\newcommand{\tablefont}{\small}
\newcommand{\sym}[1]{\rlap{$^{#1}$}}
\title{Competitive satellite placement and the economic geography of the geostationary orbit}
\author{Akhil Rao\thanks{Middlebury College and Rational Futures} \and Nikodem Szumilo\thanks{University College London and VARi Knowledge Partners}}
\date{}

\makeatletter
\@ifpackageloaded{caption}{}{\usepackage{caption}}
\AtBeginDocument{%
\ifdefined\contentsname
  \renewcommand*\contentsname{Table of contents}
\else
  \newcommand\contentsname{Table of contents}
\fi
\ifdefined\listfigurename
  \renewcommand*\listfigurename{List of Figures}
\else
  \newcommand\listfigurename{List of Figures}
\fi
\ifdefined\listtablename
  \renewcommand*\listtablename{List of Tables}
\else
  \newcommand\listtablename{List of Tables}
\fi
\ifdefined\figurename
  \renewcommand*\figurename{Figure}
\else
  \newcommand\figurename{Figure}
\fi
\ifdefined\tablename
  \renewcommand*\tablename{Table}
\else
  \newcommand\tablename{Table}
\fi
}
\@ifpackageloaded{float}{}{\usepackage{float}}
\floatstyle{ruled}
\@ifundefined{c@chapter}{\newfloat{codelisting}{h}{lop}}{\newfloat{codelisting}{h}{lop}[chapter]}
\floatname{codelisting}{Listing}

\makeatother
\makeatletter
\@ifpackageloaded{caption}{}{\usepackage{caption}}
\@ifpackageloaded{subcaption}{}{\usepackage{subcaption}}
\makeatother
\usepackage{bookmark}
\IfFileExists{xurl.sty}{\usepackage{xurl}}{} 
\hypersetup{
  pdftitle={Competitive satellite placement and the economic geography of the geostationary orbit},
  pdfauthor={Akhil Rao and Nikodem Szumilo},
  colorlinks=true,
  linkcolor={blue},
  filecolor={Maroon},
  citecolor={Blue},
  urlcolor={Blue},
  pdfcreator={LaTeX via pandoc}}

\begin{document}

\maketitle

\begin{abstract}
\noindent The geostationary orbit (GEO) carries most of the world's satellite communications revenue. The International Telecommunication Union (ITU) coordinates and records frequency assignments and associated orbital positions through national administrations, without creating a market price or property title for longitude. We ask how commercial operators locate along GEO within this administrative regime. Using a spatial sorting model, we show that commercial satellites are placed in proportion to the reachable income below a GEO longitude, discounted by viewing angle at a spatial discount rate. We estimate a placement elasticity close to one, so that satellite counts above a longitude rise proportionately with reachable income. Government satellites, by contrast, are placed in proportion to the population a GEO longitude covers. We validate the commercial model by showing that parameters estimated on 2012 placements predict the 2021 distribution well and that the model rejects treating Chinese GDP as fully addressable to the international commercial fleet. Counterfactuals suggest that competition from low-Earth orbit (LEO) systems may tilt GEO usage toward Asia-Pacific, while higher spatial discounting may concentrate it over high-GDP countries.
\end{abstract}\medskip


\section{Introduction}\label{sec:intro}

\begin{figure}[htbp]
\centering
\includegraphics[width=\textwidth]{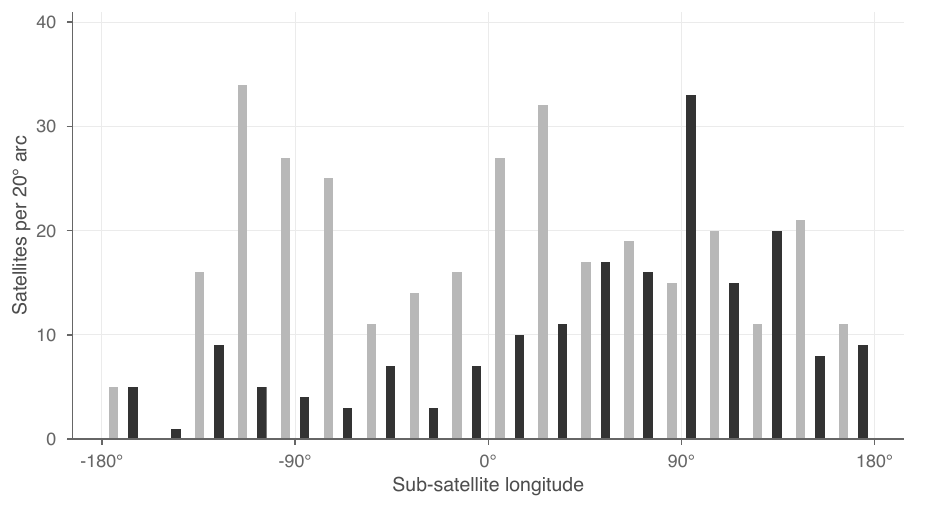}
\caption{Commercial and government satellites are geographically distinct. Satellites per $20^\circ$ longitude arc for the two sectors; the commercial belt peaks over the Americas and Asia and the government belt over Asia. The commercial series is the lighter bars and the government series the darker, as labeled on the panel axes.}
\label{fig:gov-comm}
\end{figure}

Satellite television, delivered almost entirely from geostationary orbit
(GEO), was a \$77 billion business in 2023, about \(70\%\) of all
satellite-service revenue \citep{sia2024}. GEO is thus among the most
valuable strips of real estate off the Earth's surface. It is also used
unevenly, with satellites clustering over some locations (e.g., Asia,
Europe, and the Americas), while long stretches over the open oceans sit
empty. Further, it is heavily used by both commercial and government
satellite operators, and government operators appear to systematically
choose different locations than commercial operators. This paper
develops an economic model of location choice to explain these patterns,
shown in Figure~\ref{fig:gov-comm}.

GEO is a governed common-pool resource
\citep{weeden2012common, ostrom1990governing}: it is rival, since nearby
networks can interfere and two satellites cannot occupy the same
location at the same time, but operators cannot buy longitudes on a
market \citep{macauley1998allocation}. They can only file for them
through the ITU, which allocates spectrum to services and administers
the procedures through which national administrations coordinate,
notify, and record frequency assignments together with associated
orbital positions. Recording provides international recognition against
harmful interference, not ownership of a freestanding slot
\citep{itu-rr-2024, itu-constitution-2019}. National administrations
decide which operators may use the resulting assignments, through
administrative, comparative, or market-based licensing. How do operators
sort across the belt under this management structure, absent an
international market for tradable orbital locations?

We focus on commercial operators, who operate satellites above
particular longitudes to sell telecommunications services.
Telecommunications signals degrade with greater distance between sender
and receiver, so locations farther from the longitude directly under a
slot may experience lower quality signals than locations closer to
directly under a slot. We assume consumers prefer higher quality
signals, so that a satellite receives a larger share of demand from
consumers directly under it than from consumers farther away. We build a
spatial sorting model in which satellite operators choose GEO slots
based on the demand they may capture from that slot. Operators enter a
slot as long as the returns from the slot cover the cost of occupying
it, and sort across slots until returns are equal everywhere (i.e.,
``free-entry''). The equilibrium condition of this sorting process gives
us our estimation equation. We use the same equation to study the
determinants of government satellite location choice.

Three findings follow. First, commercial satellites' location choices
are consistent with a free-entry model in which operators prefer
longitudes with better views of higher-GDP areas. Parameters estimated
on 2012 placements predict 2021 placements well. As a further
validation, we use the fact that Chinese demand is not addressable to
foreign competitors, and test whether treating Chinese GDP as fully
addressable improves fit. We find it does not. Second, we estimate the
placement elasticity, i.e., the elasticity of a longitude's occupancy
with respect to the demand it can reach. A placement elasticity of one
means satellite counts rise proportionately with demand. We find that
the placement elasticity is not distinguishable from one. Finally,
government satellite placements are consistent with governments aiming
to maximize population covered rather than maximize commercial revenue.

We then consider two counterfactual scenarios. The first involves
low-Earth orbit (LEO) telecommunications constellations, such as
Starlink \citep{spacex-s1-2026, farrar2025}. We find that if LEO
constellations capture demand from higher-GDP regions, commercial GEO
satellites will eventually shift to cover lower-GDP regions. The second
involves increasing the consumer preference for higher-quality signals,
consistent with improved terrestrial substitute options. We find that
this will lead commercial GEO satellites to concentrate over higher-GDP
regions.

This paper contributes to a growing literature on the economics of
orbital resources. While an earlier strand assessed orbit-spectrum
allocation in GEO \citep{macauley1998allocation}, more recent work has
focused on debris externalities and market concentration in LEO
\citep{adilov2015economic, wagner-grzelka-2019, rao2020orbital, guyot2023oligopoly, rao-rondina-2025, doyle2026}.
Our focus is on how the geography of demand and market access shapes GEO
location choice. \citet{kaffine-rao-2026} is the most closely related
analysis, asking whether observed orbital use in LEO is consistent with
open-access, profit-maximizing entry subject to collision risk. We
instead study placement within a coordinated administrative regime and
estimate the the response of occupancy to reachable income in a model of
profit-maximizing sorting. Such sorting across markets or spatially
distributed resources has been studied in settings ranging from cities,
traffic, recreational fishing sites, and abstract geometries
\citep{harris1954market, salop1979monopolistic, bayer-timmins-2007, timmins-murdock-2007, glaeser2008cities, hughes-kaffine-2017}.

\section{Institutional background and the physics of the belt}\label{sec:background}

\subsection{The geostationary belt as a physical resource}\label{sec:physics}

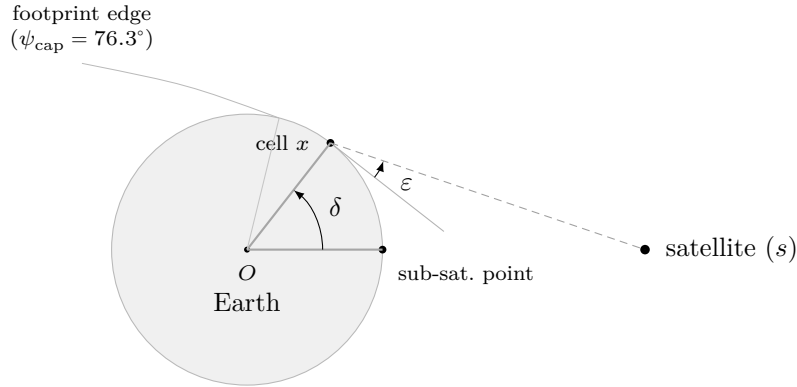
\begin{figure}[htbp]
\centering
\begin{tikzpicture}[scale=1.12,>=latex,font=\small]
  \def\Re{1.6}      
  \def\rgeo{4.7}    
  \coordinate (O) at (0,0);
  \draw[fill=gray!12,draw=gray!55] (O) circle (\Re);
  \node at (0,-0.62) {\small Earth};
  \fill (O) circle (1pt); \node[below=2pt] at (O) {\scriptsize $O$};
  \coordinate (S) at (\rgeo,0);
  \fill (S) circle (1.6pt); \node[right=4pt] at (S) {\small satellite ($s$)};
  \coordinate (P) at (\Re,0);   
  \fill (P) circle (1.3pt); \node[below right=2pt] at (P) {\scriptsize sub-sat.\ point};
  \def\dlt{52}
  \coordinate (X) at ({\Re*cos(\dlt)},{\Re*sin(\dlt)});
  \fill (X) circle (1.3pt); \node[left=4pt] at (X) {\scriptsize cell $x$};
  \draw[thick,gray!70] (O) -- (P); \draw[thick,gray!70] (O) -- (X);
  \pic[draw,->,"$\delta$",angle radius=1.0cm,angle eccentricity=1.3]{angle=P--O--X};
  \draw[densely dashed,gray!75] (X) -- (S);
  \coordinate (H) at ($(X)!1.7cm!90:(O)$);   
  \draw[gray!60] (X) -- (H);
  \pic[draw,->,"$\varepsilon$",angle radius=0.74cm,angle eccentricity=1.55]{angle=H--X--S};
  \def\psicap{76.3}
  \coordinate (E) at ({\Re*cos(\psicap)},{\Re*sin(\psicap)});
  \draw[gray!45] (O) -- (E);
  \node[anchor=south,align=center,font=\scriptsize] (FL) at (-1.95,2.2)
       {footprint edge\\($\psi_{\text{cap}}=76.3^\circ$)};
  \draw[gray!55,line width=0.3pt] (FL.south) .. controls (-0.7,1.95) .. (E);
\end{tikzpicture}
\caption{Viewing geometry of a geostationary slot. A satellite at orbital longitude $s$ sees
a terrestrial cell $x$ at Earth-central angle $\delta$ (measured at the Earth's center $O$
from the sub-satellite point) and is seen from $x$ at elevation angle $\varepsilon$ above the
local horizon. The cell is inside the footprint while $\delta\le\psi_{\text{cap}}=76.3^\circ$,
the half-angle at which $\varepsilon$ falls to the $5^\circ$ minimum for reliable service.
Equations~\eqref{eq:delta} and \eqref{eq:elevation} make these angles precise.}
\label{fig:geometry}
\end{figure}

A geostationary satellite orbits at about \(35{,}786\) km above the
equator, where its period matches the Earth's rotation, so it hovers
over a fixed longitude and a ground antenna can point at it without
tracking. Since a satellite is identified by the longitude it holds, GEO
is effectively a one-dimensional resource. A terrestrial cell at
latitude \(\phi_x\), longitude \(\theta_x\) is viewed from a satellite
at orbital longitude \(s\) at Earth-central angle \begin{equation}
\delta(x,s)=\arccos\!\big(\cos\phi_x\,\cos(\theta_x-s)\big),
\label{eq:delta}
\end{equation} and lies in the footprint when
\(\delta(x,s)\le\psi_{\text{cap}}=76.3^\circ\), the half-angle at a
\(5^\circ\) minimum elevation, a standard service floor below which
atmospheric attenuation rises sharply \citep{itur-p618}.
Figure~\ref{fig:geometry} shows the geometry: the Earth-central angle
\(\delta\), the elevation angle \(\varepsilon\) at which the satellite
clears a user's horizon, and the footprint edge where \(\varepsilon\)
falls to the minimum a ground antenna needs. Demand farther from the
sub-satellite point \(s\) is therefore worth less to an operator at
\(s\). The viewing angle \(\delta(x,s)\) is scaled by the inverse of a
spatial discount rate \(\lambda\), which sets the angular scale of
reachable demand. We describe the estimation in more detail in Section
\ref{sec:model}.

\subsection{ITU coordination and the orbital commons}\label{sec:institutions}

Orbital space cannot be appropriated as national territory under the
1967 Outer Space Treaty \citep{outer-space-treaty-1967}, so a national
administration does not acquire title to a longitude. Because the
international procedure has no auction or posted price for a longitude,
GEO is a governed common-pool resource rather than private property or
unmanaged open access
\citep{macauley1998allocation, weeden2012common, ostrom1990governing}.
Under its Constitution and Radio Regulations, the ITU allocates
frequency bands to services and administers the procedures through which
national administrations coordinate, notify, and record frequency
assignments together with associated orbital positions
\citep{itu-constitution-2019, itu-rr-2024}. National administrations
decide which operators may use the resulting assignments. They may
auction particular frequency bands to firms, but in doing so they
control access to their national market and the use of those bands to
reach it. Recent work proposes using that control as a policy
instrument, conditioning access to national markets and facilities on
operator compliance by analogy to maritime port-state control
\citep{rieder-wagner-2025, abuelenin2025port}; the same lever is studied
beyond space in trade and environmental law
\citep{scott2014extraterritoriality}.

Government use of GEO is extensive. Civil and military satellites make
up a large share of the belt (Section \ref{sec:data}), so commercial and
government placement are separate choice problems. The line between
state and commercial use is also not sharp. Intelsat began in 1964 as an
intergovernmental consortium under the 1962 COMSAT Act and was
privatized only in 2000
\citep{comsat-act-1962, orbit-act-2000, gao2004intelsat}. Equatorial
states pressed a sovereignty claim over the arc above their territory in
the 1976 Bogotá Declaration \citep{bogota-declaration-1976}; the claim
did not become the governing rule, and the present system rests instead
on non-appropriation, national filings, international coordination, and
recording.

\section{A model of commercial placement}\label{sec:model}

\subsection{The commercial operator's choice problem}\label{sec:choice}

A commercial operator chooses a longitude to earn revenue from the
terrestrial demand below it. Index demand by cell \(x\) with latitude
\(\phi_x\), longitude \(\theta_x\), and demand \(D(x)\), where \(D\) is
reachable GDP for the commercial problem (Section~\ref{sec:data}). A
satellite at orbital longitude \(s\) views cell \(x\) at the
Earth-central angle \(\delta(x,s)\) of Equation \eqref{eq:delta}, and
the cell is inside the footprint when
\(\delta(x,s)\le\psi_{\text{cap}}=76.3^\circ\). Demand farther from the
sub-satellite point is worth less, because slant range and atmospheric
path lengthen as elevation falls and signal quality degrades with them.

How much less depends on consumer behavior. We assume a consumer
tolerates some degradation before dropping the service, and we call the
viewing angle at which a consumer stops subscribing their angular
tolerance. Suppose the only thing we know about the distribution of
tolerances in the population is its mean, \(\lambda\) degrees. The
distribution that is consistent with that mean and assumes nothing
further is exponential with rate \(\lambda^{-1}\). Take-up in a cell is
the share of consumers whose tolerance exceeds the angle at which the
satellite serves them, which is the survival function of that
distribution: \begin{equation}
\sigma(x,s)=\rho\,\exp\!\big(-\delta(x,s)/\lambda\big),
\label{eq:takeup}
\end{equation} where \(\rho\) is take-up directly beneath the satellite.
A cell contributes its income times its take-up, so the revenue a slot
reaches is \(\sum_x D(x)\,\sigma(x,s)=\rho\,\mathrm{ED}(s;\lambda)\),
where \begin{equation}
\mathrm{ED}(s;\lambda)=\sum_{x:\,\delta(x,s)\le\psi_{\text{cap}}}
  D(x)\,\exp\!\big(-\delta(x,s)/\lambda\big)
\label{eq:ma}
\end{equation} is the slot's effective demand, the footprint demand
discounted by viewing
angle.\footnote{While we motivate the exponential kernel by a maximum entropy principle \citep{jaynes-1957}, the same kernel arises from a random-utility model in which a consumer with value $v$ net of price, disutility $\lambda^{-1}$ per degree of viewing angle, and an exponentially distributed idiosyncratic taste draw subscribes whenever net utility is positive. Take-up is then $\exp(v-\delta/\lambda)$, with $\rho=e^{v}$. This is a binary response model with a log link: $\lambda^{-1}$ is a constant multiplicative effect on the take-up probability rather than on the odds \citep{donoghoe-marschner-2018}. Such a model requires the linear predictor to be negative so that take-up cannot exceed one \citep{donoghoe-marschner-2018}, which here says only that satellite service is a minority good even directly overhead. Any taste distribution with an exponential upper tail delivers the same kernel in the region where take-up is small.}

The reach \(\lambda\) sets the angular scale of the discount: a cell
\(\delta\) degrees off the sub-satellite point enters with weight
\(e^{-\delta/\lambda}\), so a slot's effective demand falls at the rate
\(\lambda^{-1}\) per degree of viewing angle. We call \(\lambda^{-1}\)
the slot's spatial discount rate; at the estimated reach it is about
\(6\%\) per degree, the angular analogue of the per-year discounting of
distant benefits in resource economics \citep{yamaguchi-shaw-2020}. With
a wide reach (low spatial discount rate), effective demand approaches
the undiscounted footprint total; with a narrow reach (high spatial
discount rate), the weight concentrates near the sub-satellite point.

The value of a longitude is proportional to the revenue an operator can
earn there. Gross reachable revenue at \(s\) is
\(R(s)=\rho\,\mathrm{ED}(s;\lambda)\), where \(\rho\) converts a unit of
effective demand into dollars. Longitudes are not traded at a price, so
we never observe an operator's willingness to pay for one. If operators'
willingness to pay is proportional to a longitude's profitability, and
if the cost of entering is about equal everywhere, then profitability is
proportional to reachable revenue \(\rho\,\mathrm{ED}(s;\lambda)\). We
therefore treat effective demand as an unpriced amenity index in a
spatial sorting model and infer a longitude's value from where operators
place satellites \citep{timmins-murdock-2007, bayer-timmins-2007}.
Multiple satellites may operate within the same one-degree bin and
contest overlapping customers.

We represent the resulting local crowding by allowing the return to any
one operator to fall as the count \(N(s)\) rises: \begin{equation}
r(s)=\frac{\rho\,\mathrm{ED}(s;\lambda)}{N(s)^{1/\beta}}.
\label{eq:per-operator}
\end{equation} The exponent \(1/\beta>0\) makes per-operator returns
fall as a longitude fills: in logs,
\(\log r(s)=\log\rho+\log\mathrm{ED}(s;\lambda)-(1/\beta)\log N(s)\).
This declining return is the orbital counterpart of local crowding in
models of sorting across a common-pool resource
\citep{bayer-timmins-2007, timmins-murdock-2007, hughes-kaffine-2017}.

We assume each operator chooses the longitude it values most. Operators
are not identical, but most of the ways they differ do not affect this
choice. An operator's overall cost, scale, or efficiency shifts the
value of every longitude alike and so cancels from any comparison
between two longitudes. Characteristics of the longitude itself do
affect the choice. Unobserved factors relevant to a slot's value include
regulatory access to the markets it faces, the terrestrial services it
would compete with, and economic conditions not reflected in national
income. We collect these into a slot-level demand shock \(\eta_s>0\),
common to every operator at that longitude, so operator \(j\) solves the
location problem \begin{equation}
s_j^\ast\ \in\ \arg\max_{s}\ \Big\{\,\log\rho+\log\mathrm{ED}(s;\lambda)+\log\eta_s
  -\tfrac{1}{\beta}\log N(s)\,\Big\},
\label{eq:choice}
\end{equation} choosing the slot with the highest return. Because
\(\eta_s\) is common to the operators at a slot, it moves what a
longitude is worth without disturbing how operators rank longitudes.
Operators best-respond to the occupancy they anticipate: the count
\(N(s)\) each conditions on is a slot's expected occupancy, which
equilibrium requires to equal the count that operators' own choices
generate, so expectations are validated in equilibrium.

Aggregating these choices pins down the equilibrium. Because every
operator solves Equation~\eqref{eq:choice}, entry flows toward any
longitude offering an above-average return until no relocation pays, so
in equilibrium the per-operator return is equalized across all occupied
longitudes at a common level \(\bar r\) that just covers the cost of
entry: \begin{equation}
r(s)=\bar r \qquad\text{for every occupied } s.
\label{eq:equalization}
\end{equation}

The estimating equation follows from the equilibrium condition. Setting
\(r(s)=\bar r\) in Equation~\eqref{eq:per-operator} and rearranging
gives
\(N(s)=(\rho\,\eta_s/\bar r)^{\beta}\,\mathrm{ED}(s;\lambda)^{\beta}\):
the occupied count is a constant elasticity \(\beta\) in effective
demand, scaled by the slot's unobserved demand shock. Averaging over
that shock, which we take to be mean-independent of effective demand,
gives the equation we take to the data, \begin{equation}
\mathbb{E}[N(s)]=\exp\!\big(\alpha+\beta\log\mathrm{ED}(s;\lambda)\big),
\label{eq:intensity}
\end{equation} with intercept \begin{equation}
\alpha=\beta\log(\rho/\bar r)+\log\mathbb{E}\big[\eta_s^{\beta}\big].
\label{eq:alpha}
\end{equation} When \(\beta=1\), equilibrium occupancy is proportional
to effective demand, and the fixed reachable-revenue pool is divided
evenly among the operators at a longitude. When \(\beta<1\), occupancy
responds less than proportionally to effective demand; when \(\beta>1\),
more than proportionally. We estimate Equation \eqref{eq:intensity} by
Poisson pseudo-maximum-likelihood over all \(360\) one-degree slots,
which is a consistent estimator as long as the conditional mean is
correctly specified \citep{santos-silva-tenreyro-2006}. Because Equation
\eqref{eq:per-operator} imposes a strictly declining return for every
finite \(\beta>0\), the equilibrium described by Equation
\eqref{eq:intensity} exists and is unique.

The remaining primitive is the parameter governing which longitudes are
worth occupying at all. A longitude draws at least one operator only if
its reachable revenue covers the fixed entry cost \(F\), that is
\(R(s)=\rho\,\mathrm{ED}(s;\lambda)\ge F\), or \begin{equation}
\mathrm{ED}(s;\lambda)\ \ge\ \tau\equiv F/\rho.
\label{eq:entry}
\end{equation} We calibrate the threshold \(\tau\) to observed
occupancy, setting it to the effective demand of the least-accessible
occupied longitude; it summarizes how high effective demand must run for
a longitude to be worth occupying. Section~\ref{sec:commercial} reports
the value, though the Section~\ref{sec:counterfactuals} counterfactuals
perturb the demand surface without re-applying \(\tau\). Appendix
\ref{app:methods} describes the mechanics in more detail.

\subsection{Data}\label{sec:data}

Satellite longitudes are from the ITU Compliance Assessment Monitor
\citep{roberts2024method}, with each satellite mapped to its nearest
one-degree slot; operator role (commercial, civil, military) is from the
AEI operational-satellite table \citep{aei-spacedata}, joined on NORAD
identifier. For each year, we construct the GEO location distribution
using satellite locations on August 1. Appendix~\ref{app:data} provides
more details.

The unit of analysis is the one-degree longitude bin and the outcome is
\(N(s)\), the number of operational satellites stationed at longitude
\(s\). We take the longitude, rather than the satellite or the operator,
as the unit because a longitude's value and average occupancy are stable
while satellite identities churn with retirement and replacement. We
count only satellites operational at the snapshot date; those boosted to
a graveyard orbit at end of life are supersynchronous and drift across
all longitudes, so dropping them keeps disposal out of the occupancy
measure \citep{roberts2024thesis}. Both government and commercial
satellites occupy well under half the 360 one-degree longitudes (Table
\ref{tab:summary}) and place a large share of their satellites in a few
arcs (Figure \ref{fig:gov-comm}).

\begin{table}[ht]
\centering
{\tablefont
\begin{tabular}{l rrr}
\toprule
 & \multicolumn{1}{c}{Satellites} & \multicolumn{1}{c}{Occupied} & \multicolumn{1}{c}{Satellites per} \\
 & & \multicolumn{1}{c}{longitudes} & \multicolumn{1}{c}{occupied long.} \\
\midrule
Commercial & 321 & 189 & 1.70 \\
Government & 183 & 130 & 1.41 \\
\bottomrule
\end{tabular}
}
\caption{Summary statistics for GEO in 2021. For each user type: the
number of satellites, how many of the 360 one-degree longitudes it occupies, and the mean number of satellites per occupied longitude.}
\label{tab:summary}
\end{table}

\begin{figure}[htbp]
\centering
\includegraphics[width=0.84\textwidth]{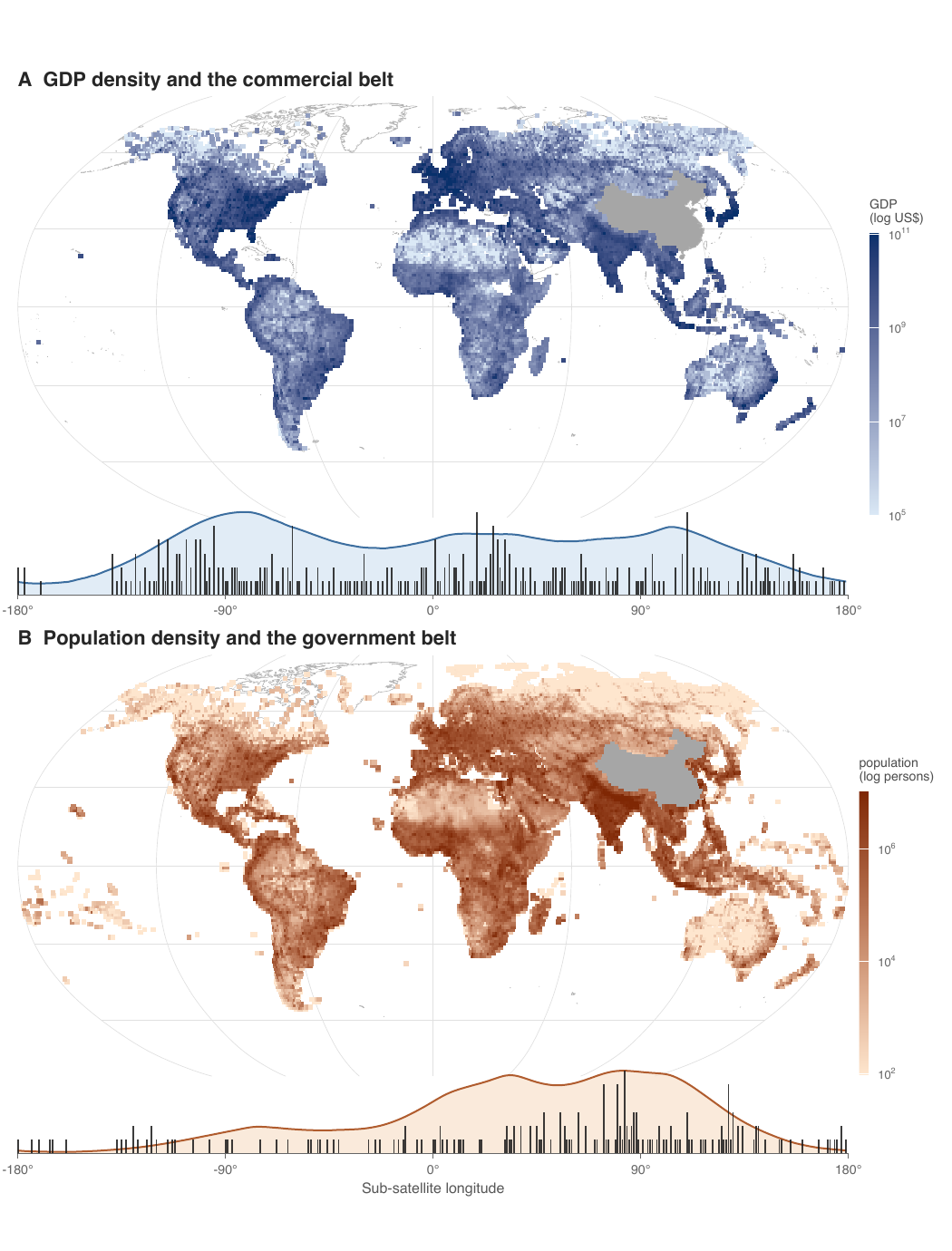}
\caption{Each panel maps a
demand surface on a log scale, with the belt below it as a longitude
profile: effective demand at each orbital longitude (filled), with the satellite belt
overlaid as vertical spikes; each profile series is normalized to its own maximum.
\textbf{A}: GDP density and the commercial belt. \textbf{B}: population density and the
government belt. Commercial satellites cluster over the GDP peaks (the Americas and Asia);
government satellites over the population peak (Asia). Annual market-exchange-rate GDP,
2021; China excluded from the demand surfaces as in the commercial model estimation.}
\label{fig:demand-panels}
\end{figure}

Figure \ref{fig:demand-panels} overlays the commercial satellites on
reachable GDP, and the government satellites on reachable population.

Demand surfaces are built on one-degree annual grids: reachable GDP for
the commercial problem, from World Bank national accounts distributed
over within-country spatial shares of a gridded GDP product
\citep{worldbank-wdi, kummu2025gdp}, and reachable population for the
government problem, from WorldPop \citep{lloyd2017worldpop}. GDP is
recorded annually, while population is linearly interpolated or
extrapolated from estimates for 2010, 2015, and 2020. We exclude China
from the commercial demand surface because Chinese end users are not
directly contestable by the international commercial operators in our
sample. China's telecommunications regulator classifies satellite
communication service, and the business of leasing transponder capacity
to users inside China, as basic telecommunications services
\citep{miit-2015-telecom-catalogue, prc-telecom-regulations-2000}.
Foreign equity in a basic-telecom enterprise is capped at 49 percent by
statute \citep{prc-order333-2001-foreign-invested-telecom}, and in
practice no foreign-invested basic-telecom enterprise has been approved
\citep{birdbird-2022-china-fite-reform}. Commercial access to the
Chinese market runs instead through wholesale capacity arrangements with
domestic licensees, and the operators that dominate capacity over the
China-facing arc (\(\sim\)\relax75--140\(^\circ\)E) are state-controlled
or state-affiliated: China Satcom is a subsidiary of the state-owned
China Aerospace Science and Technology Corporation, APT Satellite is
majority controlled by the same group, and AsiaSat is jointly controlled
by the state-owned CITIC and a private-equity partner. Rents over that
arc accrue to these domestic operators rather than to international
entrants. The exclusion applies only to the demand surface; slots in
that longitude range remain in the model and still attract South and
Southeast Asian demand. Appendix~\ref{app:data} details construction.

\subsection{Commercial placement}\label{sec:commercial}

\begin{figure}[htbp]
\centering
\includegraphics[width=\textwidth]{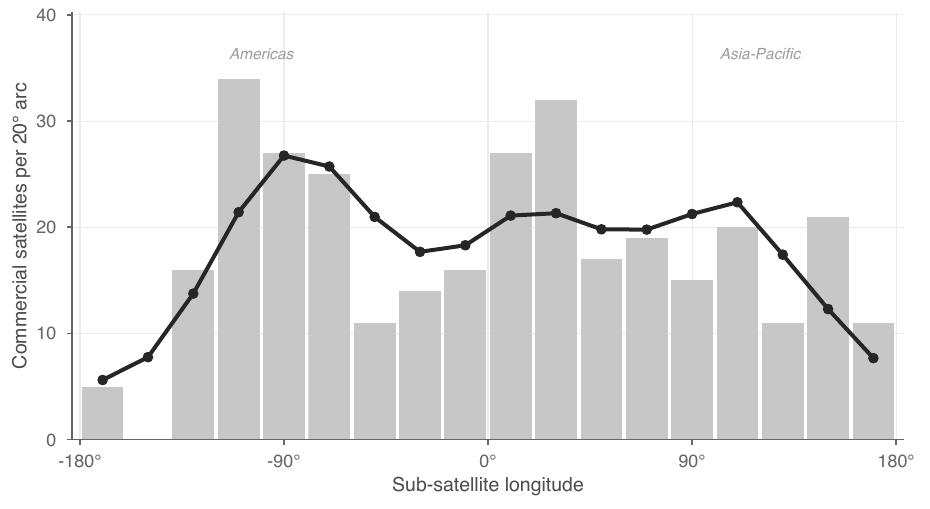}
\caption{The model tracks the commercial belt. Commercial satellites per $20^\circ$
longitude arc: observed counts (bars) against the Poisson-predicted counts from reachable GDP
(connected points). The model reproduces the belt's shape, dense over the Americas and Asia and
thin over the oceans, and is smoother than the observed counts. Counts are aggregated to
$20^\circ$ arcs for legibility.}
\label{fig:fit-commercial}
\end{figure}

An operator earns revenue from the demand its longitude reaches, so we
expect commercial counts to rise with reachable GDP rather than with
reachable population. Table~\ref{tab:diss}, columns (1)-(3), show that
this holds. In univariate regressions, the model with reachable GDP fits
the commercial satellite counts better (i.e., higher maximized
log-likelihood) than the model with population. In the bivariate
regression, predicted commercial counts rise with reachable GDP and are
flat in population. The placement density the model predicts from
reachable GDP tracks the observed commercial belt across longitude
(Figure \ref{fig:fit-commercial}).

\begin{table}[ht]
\centering
{\tablefont
\setlength{\tabcolsep}{9pt}%
\begin{tabular}{l r r r @{\hskip 2em} r r r}
\toprule
 & \multicolumn{3}{c}{Commercial belt} & \multicolumn{3}{c}{Government belt} \\
\cmidrule(lr){2-4}\cmidrule(lr){5-7}
 & \multicolumn{1}{c}{(1)} & \multicolumn{1}{c}{(2)} & \multicolumn{1}{c}{(3)} & \multicolumn{1}{c}{(4)} & \multicolumn{1}{c}{(5)} & \multicolumn{1}{c}{(6)} \\
\midrule
$\log$ reachable GDP & $+0.82$\sym{***} &  & $+0.87$\sym{***} & $+1.39$\sym{***} &  & $-0.45$ \\
 & $(0.20)$ &  & $(0.25)$ & $(0.46)$ &  & $(0.39)$ \\
$\log$ reachable population &  & $+0.24$\sym{**} & $-0.03$ &  & $+0.89$\sym{***} & $+1.00$\sym{***} \\
 &  & $(0.11)$ & $(0.09)$ &  & $(0.18)$ & $(0.23)$ \\
\midrule
Longitude bins & 360 & 360 & 360 & 360 & 360 & 360 \\
Satellites & 321 & 321 & 321 & 183 & 183 & 183 \\
Log-likelihood & $-456$ & $-467$ & $-456$ & $-345$ & $-327$ & $-327$ \\
\bottomrule
\end{tabular}
}
\caption{Poisson pseudo-maximum-likelihood regressions of the per-longitude satellite count on log
effective demand. The two sectors are estimated and reported in parallel: each sector's
three columns enter the demand objects progressively, GDP alone, population alone, then both.
Commercial demand is reachable GDP discounted by viewing angle (effective demand) at the
best-fit reach $\hat\lambda=17^\circ$; government demand is the full-footprint object (the
$\lambda\to\infty$ limit, uniform weight inside the $\pm76.3^\circ$ footprint). With both
objects in, commercial placement loads on
GDP and not population, and government on population and not GDP. Estimation
sample: the 2021 operational belt, 321 commercial and 183 government
satellites, with one observation per one-degree longitude bin (360 bins). The pooled-panel
elasticity reported in Section~\ref{sec:commercial} instead stacks all twelve annual cross-sections
(2010--2021); this table is the 2021 cross-section. Circular Conley spatial-HAC standard errors
($76^\circ$ bandwidth) in parentheses; $^{***}p<0.01$, $^{**}p<0.05$, $^{*}p<0.10$.}
\label{tab:diss}
\end{table}

We estimate the revenue \(\lambda\) by maximum likelihood, profiling
over \(\lambda\) with \(\alpha,\beta\) free in
Equation~\eqref{eq:intensity}. The likelihood peaks at
\(\hat\lambda=17^\circ\) (95\% confidence interval
\(11\)--\(23^\circ\)). The reach is wide relative to a consumer dish's
half-power beamwidth (\(\sim\)\relax2--3\(^\circ\) at Ku band;
\citealp{antenna-hpbw}), consistent with \(\lambda\) measuring the scale
of a served market rather than a single beam. The placement elasticity
is not distinguishable from the proportional-sharing benchmark of one.
In the cross-section \(\hat\beta=0.82\); pooling the twelve annual
cross-sections (2010--2021) with year fixed effects gives
\(\hat\beta=0.91\) with a two-way (spatial Conley and serial
slot-cluster) standard error of \(0.14\).

We set the entry threshold \(\tau\) in Equation~\eqref{eq:entry} so that
the number of occupied longitudes matches the 2021 operational
commercial satellite distribution, with 189 of 360 one-degree slots
occupied by 321 satellites, which pins \(\tau=F/\rho\) to current
occupancy without identifying \(F\) and \(\rho\) separately. The
threshold is set to \(0.65\) of the effective demand at the most
accessible longitude, i.e., a longitude is worth occupying once its
reachable market is at least about two-thirds of the belt's peak.

\begin{figure}[htbp]
\centering
\includegraphics[width=0.86\textwidth]{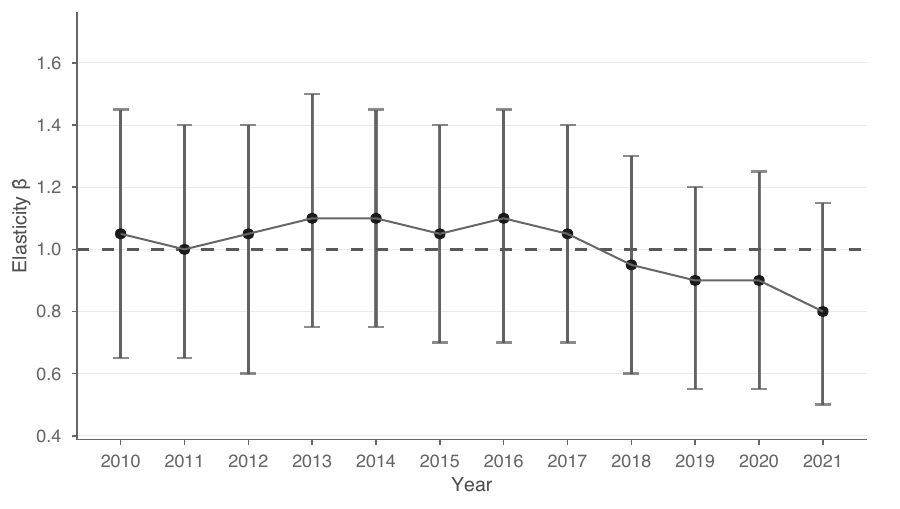}
\caption{Year-by-year estimates of the placement elasticity $\hat\beta_t$ with 95\% confidence
intervals; the reference line is the benchmark $\beta=1$.}
\label{fig:beta-t}
\end{figure}

The elasticity is stable over the sample period. Estimating
Equation~\eqref{eq:intensity} year by year on commercial satellite
placements over 2010--2021, \(\hat\beta_t\) lies within sampling error
of one in every year (Figure~\ref{fig:beta-t}). The point estimates
decline mildly, from about \(1.05\) in 2010 to \(0.80\) in 2021. The
decline could reflect drift in effective demand as a proxy for
commercial revenue, or changing operator-longitude shock dispersion,
though neither is identified under our model.

\subsection{Government placement}\label{sec:government}

\begin{figure}[htbp]
\centering
\includegraphics[width=\textwidth]{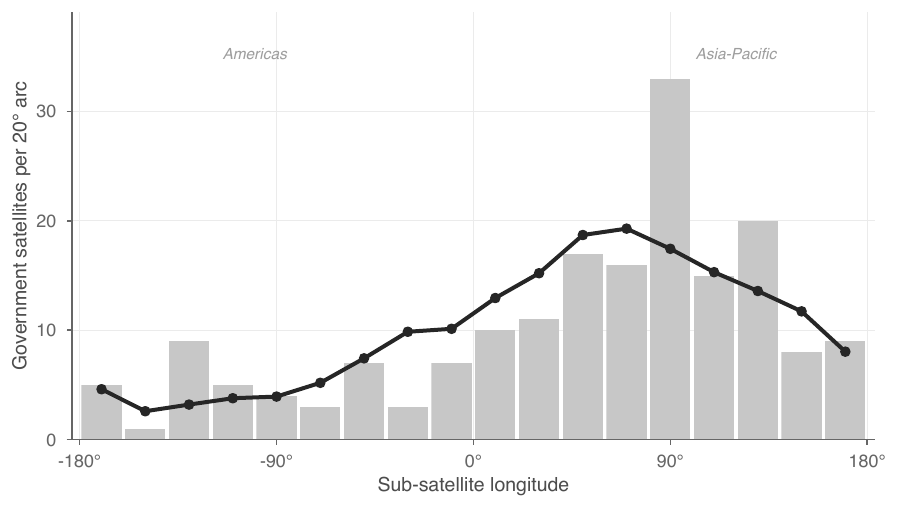}
\caption{Government (civil and military) satellites per $20^\circ$ longitude arc: observed
counts (bars) against the Poisson-predicted counts from full-footprint reachable population
(connected points).}
\label{fig:fit-government}
\end{figure}

Governments may solve a more complicated problem than commercial
operators, potentially balancing multiple objectives and competing with
other governments. Rather than derive the government location choice
equilibrium from first principles as in the commercial case, we take a
reduced-form approach and run the same Poisson specification as for the
consumer problem descriptively to see which demand object (GDP or
population) better explains government satellite placement. We use the
full-footprint objects (i.e., the \(\lambda\to\infty\) limit of
effective demand in
Equation~\eqref{eq:ma}).\footnote{We also estimated the model with $\lambda$ free, and found the likelihood surface flat across the range of possible values with slight improvement toward larger $\lambda$.}
Unlike the commercial surface, the government demand objects include
China, since a government covers its own population regardless of
whether international operators can sell there. Government counts rise
with reachable population, a pattern consistent with governments placing
satellites to cover population. The population coefficient is near one,
so government counts rise roughly in proportion to reachable population,
and reachable GDP adds no further explanatory power once population is
included. In univariate regressions, reachable population fits the
government belt better than reachable GDP (Figure
\ref{fig:fit-government}). Table \ref{tab:diss} shows the results in
columns (4)-(6).

\subsection{Out-of-sample prediction}\label{sec:oos}

To test our model of commercial placements, we assess how well it
predicts placements in future years. We estimate the model on the 2012
commercial satellite locations and 2012 reachable GDP, fix the reach and
elasticity it returns (\(\hat\lambda_{2012}=17^\circ\),
\(\hat\beta_{2012}=1.03\)), and predict the 2021 locations from 2021
reachable GDP: \begin{equation}
\hat N_{2021}(s)=\exp\!\big(\hat\alpha_{2012}+\hat\beta_{2012}\,
  \log \mathrm{ED}_{2021}(s;\hat\lambda_{2012})\big),
\label{eq:oos}
\end{equation} where \(\mathrm{ED}_{2021}\) is effective demand built
from 2021 reachable GDP evaluated at the 2012 reach. Placement decisions
resolve at the level of a served market, not a one-degree slot, so we
assess the prediction across regional arcs: the Americas (\(-135^\circ\)
to \(-30^\circ\)), Europe and Africa (\(-30^\circ\) to \(40^\circ\)),
the Middle East and South Asia (\(40^\circ\) to \(90^\circ\)), East and
Southeast Asia (\(90^\circ\) to \(150^\circ\)), and the open Pacific.

\begin{figure}[htbp]
\centering
\includegraphics[width=\textwidth]{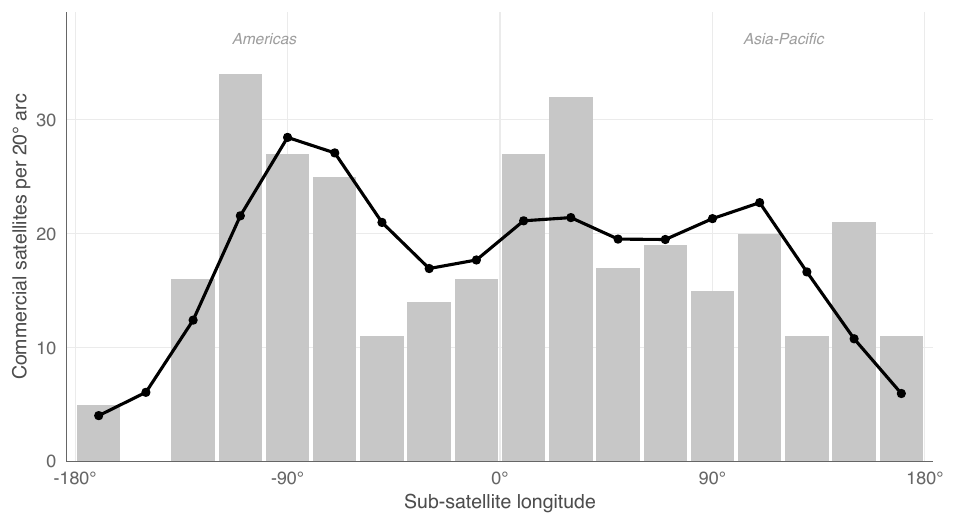}
\caption{The 2012 model predicts the 2021 belt. Commercial satellites per $20^\circ$ longitude arc
in 2021: observed counts (bars) against the counts predicted by the model estimated on the 2012 belt
and applied to 2021 reachable GDP (connected points).}
\label{fig:oos}
\end{figure}

The 2012 model reproduces 2021 location choices reasonably well (Table
\ref{tab:oos}, Figure \ref{fig:oos}). No region's predicted share of the
belt differs from its observed share by more than four percentage
points. We therefore use the model in Section \ref{sec:counterfactuals}
to study how placement would change under counterfactual demand
scenarios.

\begin{table}[ht]
\centering
{\tablefont
\begin{tabular}{l r r r r}
\toprule
 & \multicolumn{1}{c}{Observed} & \multicolumn{1}{c}{Predicted} & \multicolumn{1}{c}{Obs.\ share} & \multicolumn{1}{c}{Pred.\ share} \\
\midrule
Americas & 116 & 117 & 36.1\% & 37.2\% \\
Europe \& Africa & 82 & 69 & 25.6\% & 21.8\% \\
Middle East \& S.Asia & 42 & 49 & 13.1\% & 15.7\% \\
E. \& S.E. Asia & 51 & 57 & 15.9\% & 18.0\% \\
Pacific & 30 & 23 & 9.3\% & 7.3\% \\
\midrule
Total & 321 & 314 & & \\
\bottomrule
\end{tabular}
}
\caption{The model is
estimated on the 2012 belt, and its coefficients are applied to 2021 effective demand to predict the 2021
belt by region. Regions are the five arcs of orbital longitude defined in the text. Predicted counts are the model's Poisson means summed within each arc; the
regional correlation between predicted and observed counts is $0.97$.}
\label{tab:oos}
\end{table}

\subsection{The addressability of Chinese demand}\label{sec:china}

\begin{figure}[htbp]
\centering
\includegraphics[width=\textwidth]{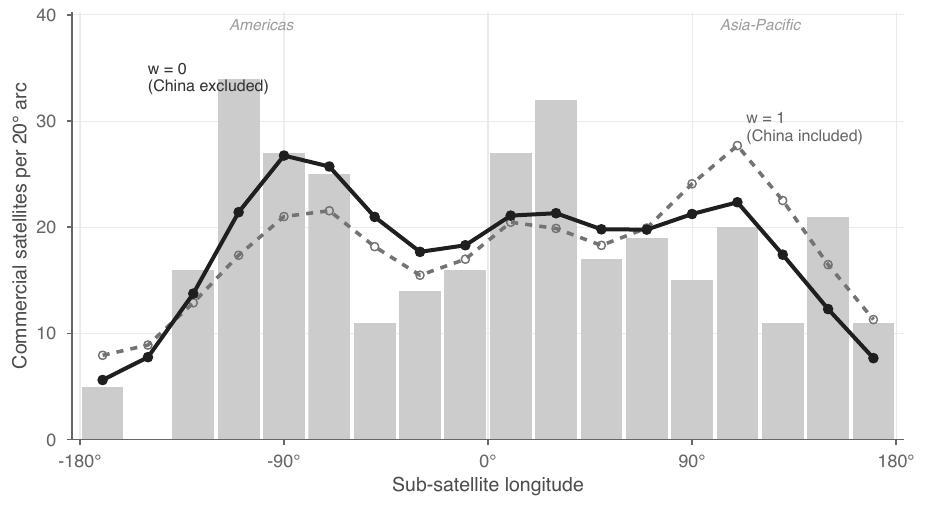}
\caption{Commercial satellites per
$20^\circ$ longitude arc: observed counts (bars) against the Poisson fitted counts under $w=0$
(China excluded, solid) and $w=1$ (Chinese demand treated as fully addressable, dashed). Both fits are at their own maximum-likelihood reach.}
\label{fig:china-fit}
\end{figure}

In our main analysis, we exclude China from reachable demand based on
institutional details that limit its addressability by international
firms. To assess whether the model supports our notion of reachability,
we test whether including Chinese GDP improves the fit. The comparison
holds the count vector and the 360 one-degree longitude bins fixed
across specifications; the only thing that changes is whether Chinese
grid cells enter effective demand. We embed that choice in a single
addressability weight \(w\), writing reachable demand as the non-China
surface plus \(w\) times the demand of Chinese cells alone, and estimate
\(w\) jointly with the elasticity \(\beta\), the reach \(\lambda\), and
the level. Excluding China is \(w=0\); treating Chinese demand as fully
addressable is \(w=1\). Table~\ref{tab:china} reports each specification
in full. The data place the weight near zero. A Wald test rejects full
addressability (\(w=1\)) at the one percent level and does not reject
zero addressability
(\(w=0\)).\footnote{A non-nested specification test confirms this is not due to linearity in $w$. The \citet{davidson-mackinnon-1981} J-test, augmenting each specification with the other's fitted values and using Conley-spatial standard errors, rejects the China-included specification at the one percent level and does not reject the China-excluded one.}
Figure~\ref{fig:china-fit} shows why full inclusion fits worse.
Including non-addressable Chinese demand concentrates predicted
placement on the \(75\)--\(140^\circ\)E arc that faces China, where the
observed commercial fleet is comparatively sparse, and reduces the
fitted counts over the Americas.

\begin{table}[ht]
\centering
{\tablefont
\setlength{\tabcolsep}{12pt}%
\begin{tabular}{l c c c}
\toprule
 & (1) China excluded & (2) China included & (3) Nested \\
 & $w=0$ & $w=1$ & $\hat w$ estimated \\
\midrule
Addressability weight $\hat w$ & $0$ & $1$ & $-0.24$ \\
\quad(Conley SE) & & & $(0.46)$ \\
\midrule
Placement elasticity $\hat\beta$ & $0.82$ & $0.28$ & $0.69$ \\
\quad(Conley SE) & $(0.20)$ & $(0.11)$ & $(0.17)$ \\
$z[\beta=1]$ & $-0.86$ & $-6.27$ & $-1.86$ \\
Reach $\hat\lambda$ (deg) & $17$ & $8$ & $14$ \\
Log-likelihood & $-456.3$ & $-463.0$ & $-455.9$ \\
\bottomrule
\end{tabular}
}
\caption{Poisson regressions of the per-longitude
commercial count on log reachable demand, estimated on the 2021 operational belt (321 satellites,
360 one-degree bins). Reachable demand is the non-China effective-demand surface plus a weight $w$ on
the demand of Chinese grid cells; column (1) imposes $w=0$ (China excluded), column (2) imposes
$w=1$ (China fully included), and column (3) estimates $w$ jointly with the elasticity, reach, and
level. Each column is at its own maximum-likelihood reach. Conley-spatial standard errors (Bartlett
kernel, $76^\circ$) in parentheses. The weight is indistinguishable from zero and full inclusion ($w=1$) is rejected at the one percent level.}
\label{tab:china}
\end{table}

\section{Counterfactuals}\label{sec:counterfactuals}

Next, we use the fitted model to study how shifts in demand would change
the geography of the belt. We run two counterfactual scenarios. In the
first, competing systems in LEO capture part of the demand a GEO
longitude can reach; in the second, each longitude's revenue reach
narrows. Both scenarios share the same structure. The baseline is the
fitted placement surface from Section \ref{sec:commercial}, the smooth
predicted distribution
\(\mu_0(s)=\exp(\hat\alpha+\hat\beta\log \mathrm{ED}(s;\lambda))\)
estimated on 2021 data. Each scenario perturbs the demand that enters
\(\mathrm{ED}(s;\lambda)\) and computes the response with the same
fitted elasticity \(\hat\beta\), so with no shock it reproduces the
fitted surface exactly. The fitted surface assigns each longitude an
expected number of satellites, and we treat a longitude as occupied with
the probability that the surface places at least one satellite there. We
report the expected number of occupied longitudes under the shock,
relative to its no-shock value and scaled to the observed 189, so the
count begins at the 2021 observed level and falls as the shock removes
effective demand.

\subsection{The effect of NGSO competition on the belt}\label{sec:ngso}

A prominent demand shock to the belt today comes from LEO. Starlink
reached \$11.4 billion in revenue in 2025 \citep{spacex-s1-2026} selling
broadband to customers a geostationary operator could potentially
otherwise serve, and it has begun disrupting incumbent geostationary
services \citep{farrar2025}. A competitor that sells the same service to
the same customers shrinks the revenue base a GEO slot can reach.
Non-geostationary (NGSO) broadband constellations in LEO are the salient
case after 2020, but the channel is general: any system that captures
the same customers reduces reachable revenue, whatever its
technology.\footnote{NGSO systems are not assigned to particular longitudes, so they can contest GEO revenue anywhere on the belt; their coexistence with geostationary networks is managed through frequency coordination and power-flux-density limits rather than any reassignment of orbital position \citep{itu-rr-2024}.}
We model the competitor as removing demand cell by cell,
\begin{equation}
D(x)\ \longrightarrow\ (1-c_x)\,D(x),
\label{eq:ngso-cut}
\end{equation} with the capture rate \(c_x\) large where the competitor
draws its customers and small elsewhere. Removing demand lowers the
effective demand of the longitudes that view those cells, so fewer
satellites are expected there and fewer longitudes remain occupied.
Because the capture is regional, the longitudes that lose the most are
those viewing the captured regions.

\begin{figure}[htbp]
\centering
\includegraphics[width=\textwidth]{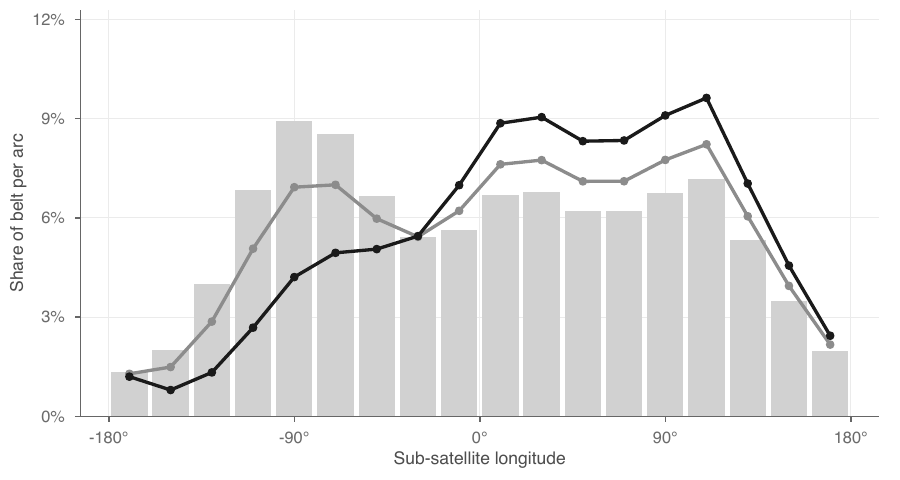}
\caption{
Share of the belt per $20^\circ$ longitude arc at the estimated reach. The bars are the
Section~\ref{sec:commercial} fitted placement surface, the no-shock baseline ($c=0$); the connected points are the predicted belt under a Starlink-traffic-weighted demand capture of $c=0.15$ and $0.30$ (darker for larger capture). The capture falls hardest on the Americas core.}
\label{fig:ngso-belt}
\end{figure}

\begin{figure}[htbp]
\centering
\includegraphics[width=0.66\textwidth]{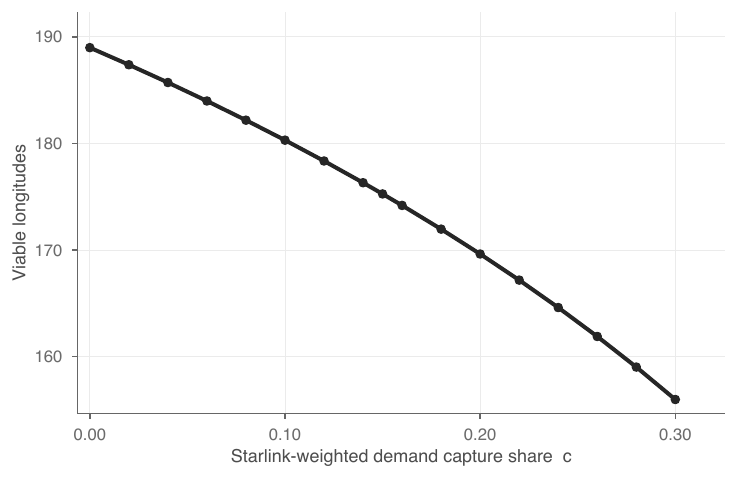}
\caption{The expected number of viable longitudes against the
aggregate Starlink-weighted demand-capture share $c$, from $0$ to $0.30$ at the estimated reach.
Viability is the model's expected count of occupied longitudes, scaled to the 189 longitudes the
commercial belt occupied in 2021, so the curve begins at 189 and falls as the capture deepens.}
\label{fig:ngso-dose}
\end{figure}

Early NGSO demand is regionally concentrated. Studies of
mega-constellation demand place most early subscribers in countries in
the upper quartile of income per person, with the Americas the densest
market \citep{rao2024demand}. We allocate the aggregate capture across
countries by an illustrative weighting concentrated on early high-income
broadband markets, so the cut falls hardest on the Americas demand core,
and compute the predicted belt (Figure \ref{fig:ngso-belt}, Figure
\ref{fig:ngso-dose}). At a 15\% aggregate capture the belt's Americas
share falls from 38\% to 32\% while its Asian share rises from 27\% to
31\%, and the expected number of viable longitudes falls from 189 to
175. At a 30\% aggregate capture the Americas share falls to 23\% and
the Asian share rises to 35\%, with 156 longitudes still viable. The
tilt away from the captured Americas core and toward the Asian arc is
robust to different capture rates.


\subsection{The effect of a higher spatial discount rate}\label{sec:lambda-cf}

The second scenario narrows each longitude's revenue reach by increasing
the spatial discount rate. Multiple mechanisms could produce a higher
spatial discount rate. For example, smaller satellites may be built to
cover a single country, trading a smaller footprint for lower cost,
making a slot's effective demand concentrate over a tighter region
\citep{astranis-microgeo}. On the demand side, rising quality
requirements make customers less willing to accept a low-elevation
signal, which raises the spatial discount rate. On the regulatory side,
tighter off-axis emission limits, set to pack slots more closely, could
have the same effect. We treat the reach reduction as an illustrative
comparative static, reducing \(\lambda\) and seeing how satellites
redistribute across GEO.

\begin{figure}[htbp]
\centering
\includegraphics[width=\textwidth]{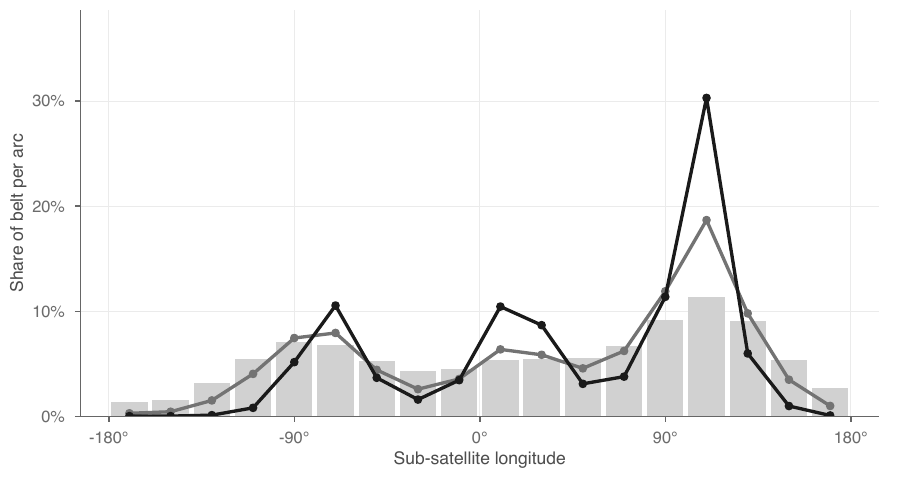}
\caption{Share of the belt per $20^\circ$
longitude arc. The bars are the fitted placement surface at the estimated reach
$\lambda=17^\circ$, the no-shock baseline; the connected points are the predicted belt at the
narrowed reaches $8.5^\circ$ and $4.25^\circ$. A higher discount rate places greater weight on demand closer to the satellite's longitude.}
\label{fig:cf-reach}
\end{figure}

A reach change acts differently from the demand shock above. Rather than
remove demand, it changes how heavily each longitude discounts the
demand it views. A narrower reach weights nearby demand more heavily, so
predicted placement concentrates over the highest-demand longitudes:
halving the reach from the estimated \(17^\circ\) to \(8.5^\circ\)
raises the Asia-Pacific share of belt mass from 37\% to 45\%, lowers the
Americas share from 32\% to 29\%, and raises the share held by the top
tenth of longitudes from 18\% to 26\% (Table \ref{tab:lambda-cf}, Figure
\ref{fig:cf-reach}).

\begin{table}[ht]
\centering
{\tablefont
\begin{tabular}{ccc}
\toprule
$\lambda$ (deg) & Americas & Asia-Pacific \\
\midrule
17 & 31.7\% & 37.1\% \\
12.75 & 31.0\% & 40.3\% \\
8.5 & 28.8\% & 45.0\% \\
4.25 & 23.6\% & 48.1\% \\
\bottomrule
\end{tabular}
}
\caption{Regional shares of
belt mass as the reach $\lambda$ narrows, computed over the Section \ref{sec:commercial} fitted placement surface (all longitudes, at the estimated elasticity $\hat\beta$, full demand geography). The fitted reach is the
angular scale over which a slot's effective demand accrues; halving it from the estimated
$17^\circ$ to $8.5^\circ$ is the headline case. Columns give the share of belt mass over the
Americas ($-150^\circ$ to $-30^\circ$) and Asia-Pacific ($60^\circ$ to $150^\circ$); the residual
Europe--Africa arc is omitted. The changes in $\lambda$ are illustrative only.}
\label{tab:lambda-cf}
\end{table}

\section{Conclusion}\label{sec:conclusion}

Commercial satellite placements appear consistent with profit-maximizing
location choice and rise approximately in proportion to the revenue a
longitude can reach. Government placements instead rise proportionately
with population. Competition from LEO broadband constellations
concentrated over the richest markets may tilt commercial placement away
from the Americas and toward Asia. A narrower revenue reach, consistent
with improved terrestrial substitutes, works in the opposite direction,
concentrating placement over its highest-demand arc.

The model is deliberately simple. We model placement one satellite at a
time, though operators manage portfolios spread across the belt, so
actual decisions are made at a coarser level. More importantly,
\(\beta\) is a reduced-form elasticity. We cannot separate the crowding
it reflects from operator-longitude match utility, inherited filings,
ground infrastructure, customer relationships, or unobserved revenue.
Appendix \ref{app:match} shows that such preference dispersion biases
\(\hat\beta\) toward zero, so the structural elasticity may be higher
than we estimate. Additional variation across bands, filing histories,
or policy regimes would be needed to identify these channels. Effective
demand is only a proxy for commercial revenue, and does not capture
geometric constraints that may differentially affect particular types of
areas, such as cities. More detailed modeling of consumers can help
address this limitation.

The belt is a governed common-pool resource whose use reflects both
economic demand and an international regulatory history. The ITU system
coordinates and records orbit-spectrum assignments, national
administrations license operators, and planned procedures preserve
formal access alongside demand-driven non-planned use, without explicit
prices or tradable rights to orbit-spectrum assignments. Despite the
lack of a formal orbital location market, commercial GEO use appears
consistent with equilibrium profit-maximizing behavior. This regularity
may help inform and assess the impacts of new governance proposals.
\newpage

\bibliography{refs}

\appendix

\section{Data}\label{app:data}
\paragraph{Satellite positions and the belt.}

Satellite longitudes are from the ITU Compliance Assessment Monitor
\citep{roberts2024method}, which records the registered and observed
positions of geostationary satellites; each satellite is mapped to its
nearest of 360 one-degree slots. For each year from 2010 to 2021 we take
the satellites operational at that year's August 1 snapshot, placed at
their contemporaneous longitudes; a satellite retired before the
snapshot is excluded from that year's belt. The reported 2021
cross-section has 321 commercial satellites at 189 occupied longitudes
and 183 government satellites (99 civil, 84 military). The 2021 belt
underlies the cross-sectional regressions, the reach estimate, and the
counterfactual entry-threshold calibration, and the full 2010--2021
panel supplies the pooled standard errors
(Section\textasciitilde{}\ref{sec:commercial}), the year-by-year
elasticities, and the out-of-sample test
(Section\textasciitilde{}\ref{sec:oos}).

\paragraph{Demand surfaces.}

The commercial demand grid is a derived annual market-exchange-rate GDP
surface for 2021: each country's World Bank national GDP (current US\$,
series NY.GDP.MKTP.CD; \citealp{worldbank-wdi}) is distributed across
one-degree land cells in proportion to the within-country spatial shares
of the Kummu et al.~gridded GDP product \citep{kummu2025gdp}. This sets
each country's cell total to its market-exchange-rate national-accounts
figure while keeping a gridded within-country pattern and, because the
World Bank series is annual, avoids any benchmark-interpolation
artifact. Reachable population is the WorldPop one-degree surface for
2021 \citep{lloyd2017worldpop}. Country labels (for the China split and
the regional capture geography) come from a Natural Earth country mask;
Chinese cells are dropped from the commercial GDP surface but retained
in the government full-footprint objects (governments cover their own
population). The government ``footprint-cap'\,' demand object in
Table~\ref{tab:diss} is the \(\lambda\to\infty\) limit of
Equation~\eqref{eq:ma}, uniform weight inside the footprint.

\section{Methods}\label{app:methods}
\paragraph{Viewing geometry.}

The Earth-central angle \(\delta(x,s)\) in Equation~\eqref{eq:delta} is
the great-circle angle between a terrestrial cell at
\((\phi_x,\theta_x)\) and the sub-satellite point \((0,s)\) of a
geostationary satellite. Placing the sub-satellite point on the equator,
the spherical law of cosines collapses to
\(\cos\delta=\cos\phi_x\cos(\theta_x-s)\). The footprint cap follows
from the minimum elevation a ground antenna needs for reliable service.
For a station at Earth radius \(R_E\) viewing a satellite at
geostationary radius \(r\), the elevation angle \(\varepsilon\) above
the local horizon and the Earth-central angle \(\delta\) satisfy
\begin{equation}
\tan\varepsilon=\frac{\cos\delta-R_E/r}{\sin\delta},
\qquad R_E/r=\frac{6378}{42\,164}\approx0.151.
\label{eq:elevation}
\end{equation} A cell is in view when \(\delta\le\psi_{\text{cap}}\);
setting \(\varepsilon=5^\circ\), a floor below which atmospheric
attenuation rises sharply \citep{itur-p618}, and solving
Equation~\eqref{eq:elevation} gives \(\psi_{\text{cap}}=76.3^\circ\).

\paragraph{Estimation.}

We estimate three objects: the elasticity (\(\beta\)), the reach
(\(\lambda\)), and the standard errors.

\textbf{The elasticity.} Equation~\eqref{eq:intensity} is fit by Poisson
pseudo-maximum-likelihood. With
\(\mu_s=\exp(\alpha+\beta\log\mathrm{ED}(s;\lambda))\), the estimator
solves the score equations \(\sum_s(N_s-\mu_s)\,w_s=0\) for the
regressors \(w_s=(1,\log\mathrm{ED}(s;\lambda))\) by iteratively
reweighted least squares over all \(360\) slots, the empty ones
included. Pseudo-maximum-likelihood is consistent under a correct
conditional mean alone \citep{santos-silva-tenreyro-2006}.
Table~\ref{tab:diss} fits each sector's count on both demand surfaces
jointly, estimated in R (\texttt{fixest}) with circular Conley
spatial-HAC standard errors at a \(76^\circ\) bandwidth (the footprint
half-width), consistent with the pooled-panel two-way errors reported in
Section~\ref{sec:commercial}.

\textbf{The reach.} The reach \(\lambda\) enters only through
\(\mathrm{ED}(s;\lambda)\), so we estimate it by profiling: for each
\(\lambda\) on a grid we concentrate out \((\alpha,\beta)\) and record
the profile log-likelihood. The maximizer is \(\hat\lambda\), and the
reported confidence interval inverts the quasi-likelihood profile with
overdispersion evaluated at \(\hat\lambda\).

\textbf{The standard errors.} For the pooled elasticity we stack the
twelve annual cross-sections (2010--2021) with year fixed effects and a
common \(\beta\). Two dependencies make the naive Poisson standard error
too small. Footprints overlap, so neighboring slots are spatially
correlated; and a given longitude's counts are serially correlated
across years. The two-way standard error combines a spatial Conley
kernel (bandwidth \(76^\circ\), the footprint half-angle, so any two
slots whose footprints overlap are treated as correlated) with a serial
slot-cluster component, computed over the \(4{,}320\) slot-years of the
stacked panel.

\paragraph{Counterfactual mechanics.}

The baseline is the fitted placement surface
\(\mu_0(s)=\exp(\hat\alpha+\hat\beta\log\mathrm{ED}(s;\lambda))\propto\mathrm{ED}(s;\lambda)^{\hat\beta}\),
the 2021 estimates (Section~\ref{sec:commercial}), positive at every
longitude. A shock is applied to the demand cells (the NGSO demand
capture) or to the kernel (reach narrowing), which moves effective
demand from \(\mathrm{ED}\) to \(\mathrm{ED}_c\); the response uses the
same elasticity,
\(\mu_c(s)=\mu_0(s)\,(\mathrm{ED}_c(s)/\mathrm{ED}(s))^{\hat\beta}\propto\mathrm{ED}_c(s)^{\hat\beta}\),
so with no shock \(\mu_c=\mu_0\) and the baseline reproduces the fitted
surface exactly. The extensive margin uses the model's own occupancy
probability. Under the fitted Poisson counts a longitude is occupied
with probability \(1-e^{-\mu(s)}\), so the expected number of occupied
longitudes is \(\sum_s\bigl(1-e^{-\mu_c(s)}\bigr)\), with the level
\(\hat\alpha\) held at its baseline value so that a demand cut lowers
the total rather than redistributing it. We report this expected count
under the shock relative to its no-shock value, scaled to the 189
longitudes occupied in 2021, so with no shock the viable count is 189 by
construction and falls as the shock removes reachable demand. The
regional shares in Figure \ref{fig:ngso-belt} show the renormalized
placement density, which describes where satellites locate given the
size of the belt, while the extensive margin uses the unnormalized
level, which describes how the size of the belt changes. The reach
counterfactual changes the kernel, not the demand level, so it has no
extensive margin and only redistributes mass. The choice of the observed
belt as the viable reference set is reported in the text; the
redistribution results do not depend on it.

\section{Operator-by-longitude match utility and the placement elasticity}\label{app:match}

The model of Section~\ref{sec:choice} lets operators differ in cost and
scale, which does not affect location choice, and lets longitudes differ
in unobserved demand, which affects every operator at a longitude alike.
It does not let a particular operator have a private reason to prefer a
particular longitude. Operators may have such reasons: a regulatory
license over the markets an arc faces, a ground network already built
there, customers and products accumulated over years. This appendix
examines the consequence of that omission: it biases the placement
elasticity toward zero.

Suppose operator \(j\) draws an idiosyncratic match utility
\(\varepsilon_{js}\) for longitude \(s\), independent across operators
and longitudes, additive on the log return of Equation
\eqref{eq:per-operator}, and distributed extreme value with scale
\(\omega\ge 0\). Operators still best-respond to anticipated occupancy.
Choice probabilities are then logit, so expected occupancy \(\nu_s\) is
proportional to \(\exp(\log r(s)/\omega)\), and with
\(x=\log\mathrm{ED}(s;\lambda)\) and \(y=\log\nu_s\) the equilibrium is
the fixed point \begin{equation}
y \;=\; \kappa + \tfrac{1}{\omega}\Big(\log\rho+\log\eta_s + x - \tfrac{1}{\beta}\,y\Big),
\label{eq:matchfp}
\end{equation} with \(\kappa\) common to all longitudes. Differentiating
Equation \eqref{eq:matchfp} with respect to \(x\) gives the elasticity
the regression recovers. Writing it
\(\tilde\beta=\mathrm{d}y/\mathrm{d}x\), \begin{equation}
\tilde\beta=\frac{1}{\omega}\Big(1-\tfrac{1}{\beta}\,\tilde\beta\Big)
\qquad\Longleftrightarrow\qquad
\frac{1}{\tilde\beta}=\frac{1}{\beta}+\omega
\qquad\Longleftrightarrow\qquad
\tilde\beta=\frac{\beta}{1+\beta\omega}.
\label{eq:attenuation}
\end{equation} Three things follow. First, the elasticity in the data is
\(\tilde\beta\), not \(\beta\): Equation \eqref{eq:intensity} is the
\(\omega=0\) case. Second, the true elasticity \(\beta\) and the match
dispersion \(\omega\) combine into the estimated
\(\tilde\beta=\beta/(1+\beta\omega)\), so the placement counts identify
only that combination. Nothing in the cross-section of occupancy
separates \(\beta\) from \(\omega\). Third, \(\tilde\beta\le\beta\) for
every \(\omega\ge 0\), with equality only at \(\omega=0\), and
\(\tilde\beta\) falls monotonically in \(\omega\): the estimated
elasticity is attenuated toward zero, and the more idiosyncratic the
operators, the more it is attenuated.

The economic content is that a match shock substitutes for measured
crowding in producing return equalization. An operator with a strong
private draw locates on a longitude even when reachable revenue is
higher elsewhere, so occupancy responds less than one-for-one to
reachable revenue. Private preferences and crowding flatten the belt
relative to the demand surface in the same way so are not separately
identified from counts.

\end{document}